\documentclass[journal]{IEEEtran}

\usepackage{tabularx}
\usepackage{cite}
\usepackage[dvips]{graphicx}
\usepackage[table, dvipsnames]{xcolor}
\usepackage{color, colortbl}
\usepackage{bbm}

\definecolor{lavenderblue}{rgb}{0.8, 0.8, 1.0}
\definecolor{skyblue}{rgb}{0.53, 0.81, 0.92}
\definecolor{blizzardblue}{rgb}{0.67, 0.9, 0.93}
	\definecolor{aqua}{rgb}{0.0, 1.0, 1.0}
   \graphicspath{{../eps/}}
   \DeclareGraphicsExtensions{.eps}

\usepackage{mathrsfs}
\usepackage{amssymb}
\usepackage{yfonts}

\usepackage[cmex10]{amsmath}
\usepackage{algorithmicx}
\usepackage{algpseudocode}

\usepackage{algorithm}
\usepackage{graphicx}
\usepackage{array}
\usepackage{color}
\usepackage[caption=false,font=footnotesize]{subfig}
\usepackage{lettrine}
\usepackage{tikz}
\usepackage{verbatim}

\newtheorem{remark}{\bf \noindent Remark}{}
\floatname{algorithm}{\textbf{Algorithm}}{}
\ifCLASSINFOpdf

\else

\fi
\begin{document}
\title{Investigating the Jamming Attack on $5$G NR Physical Channels}

\author{Mohsen~Kazemian,~\IEEEmembership{Member,~IEEE}
\thanks{M. Kazemian is with the Department of Electrical Engineering, Yazd University, Yazd, Iran, (e-mail:mohsenkazemian@yazd.ac.ir).}

}

\maketitle

\begin{abstract}
This study investigates the jamming attack on the orthogonal frequency-division multiplexing (OFDM) based physical channels in $5$G new radio (NR) technology from the aspect of signal processing. Disrupting the orthogonality property between subcarriers (SCs) is considered as one of the jammers' targets in OFDM based generations.  Focusing on the orthogonality property, we propose a method to detect the attacked subcarriers, and then neutralize the jamming attack using a multiplicative signal. Thanks to studying the frequency aspect of the attacked signal, the proposed method is independent of the jammers' transmitted power. Simulation results evaluate the detection performance of the proposed method with various numbers of OFDM subcarriers.
\end{abstract}


\IEEEpeerreviewmaketitle

\section{Introduction}
Fifth generation new radio ($5$G NR) is developed by the $3$rd generation partnership project ($3$GPP) as a radio access technology (RAT) for the $5$G cellular networks. $5$G NR operates on the higher bands between either $410$ MHz–$7125$ MHz or $24250$ MHz–$52600$ MHz. Lower latency, greater user capacity, network slicing, and enhanced speed are the superiority of this technology over the previous generations \cite{gnr}, \cite{gnrr}, \cite{gnrrr}, \cite{gnrrrr}.

Dynamic structure and removal of sparse control channels such as physical control format indicator channel (PCFICH) are two main reasons that make $5$G NR as a far less vulnerable technology compared to long term evolution (LTE) standard. Jamming attacks can only be neutralized at the physical layer, but not at the medium access control (MAC) or network layer \cite{jammm}.

In $5$G NR, uplink and downlink resources are assigned at an orthogonal frequency division multiplexing (OFDM) symbol level \cite{mine}, \cite{minee}, \cite{OFDMM}. Jammers cause an undesired denial of service (DoS) and interrupt the communications between the legitimate users and the corresponded base stations (BSs). Several scenarios can be used in this regard. The main principle to attack a channel is disrupting the orthogonally property between the OFDM spectrums by adding a phase or frequency offset to the legitimate signal. In this case, the subcarriers (SCs) experience the overlapping phenomenon and the transmitted signal will be lost entirely. The other possibilities are: $1$) transmitting radio signals with the aim of decreasing signal-to-noise ratio (SNR) at the legitimate user side, $2$) continually occupying the transmission channel to famish transmissions initiated by legitimate users, and, $3$) disabling channel estimation by forcing the received energy at the pilot OFDM samples to zero, using a fake signal similar to the legitimate one.


Physical broadcast channel (PBCH) and physical downlink control channel (PDCCH) are respectively the most vulnerable channels in $5$G NR in terms of both attack efficiency and complexity \cite{rng}.

\begin{figure}[t]
\centering
        \includegraphics[width=2.9in]{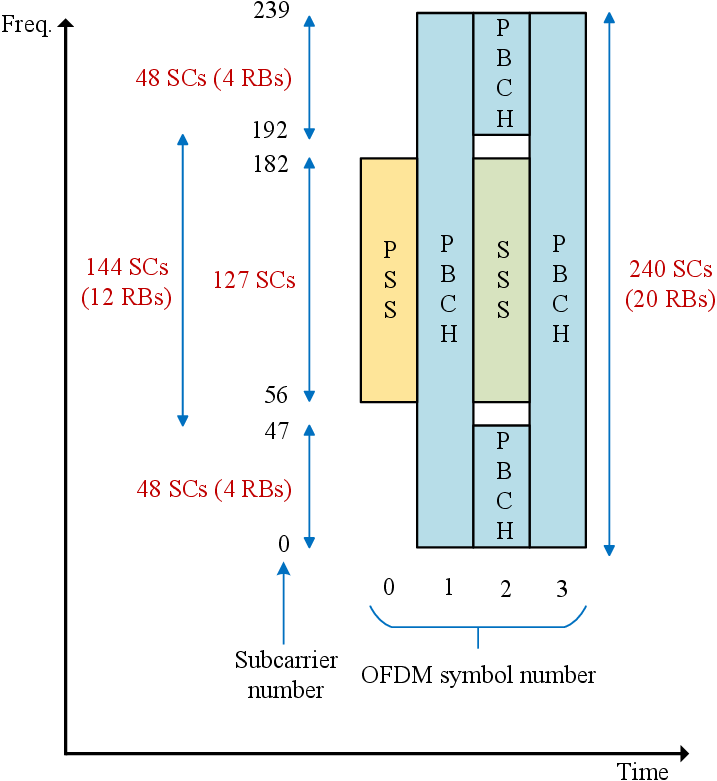}
     \caption{The time-frequency structure of SSB.}
    \label{fig:1}
\end{figure}

PBCH conveys critical information of the cell, called master information block (MIB), which includes essential information to initiate a connection between the user equipment (UE) and the corresponded cell. PBCH occupies the second and the fourth OFDM symbols (i.e., four symbols in each slot), covering $240$ subcarriers (i.e., $20$ resource blocks (RBs)) in each symbol time. Furthermore, PBCH is assigned to $48$ subcarriers, allocated to the below and the above of secondary synchronization signal (SSS) in the third OFDM symbol. Thus, PBCH occupies $576$ resource elements (REs) per synchronization signal block (SSB), including REs for PBCH payload (i.e., $432$ REs) as well as REs for demodulation reference signals (DMRS) (i.e., $144$ REs). PBCH DMRS is required for coherent demodulation of PBCH payload. The SSB structure consisting of PBCH, primary synchronization signal (PSS) and SSS is described in Fig. \ref{fig:1}.

Since unlike LTE, cell specific reference signal (CRS) is not defined in $5$G NR, PBCH DMRS is a vital signal for PBCH decoding. PBCH DMRS is considered as the most efficient target from the jammers' point of view, due to the following reasons: $1$) it is generated by pseudo-random sequence and determined by SSB, which broadcasts periodically and is unencrypted during the initial access, $2$) as it only occupies $25$\% of the PBCH resources, a malicious transmitter requires a lower level of jamming power to target it than to attack the entire PBCH block, and, $3$) PBCH DMRS is in the same spot every frame. 

After PBCH DMRS and PBCH, PDCCH ranks third in terms of vulnerability. This physical channel transmits downlink control information (DCI) as the core of the physical control channel of the $5$G network to control the transmission and reception of uplink and downlink data \cite{chan}. Physical random access channel (PRACH) and physical uplink control channel (PUCCH) are respectively placed in the next ranks. The former is used by UE to request a connection setup known as random access. PUCCH is assigned for uplink control information, including: $1$) hybrid automatic repeat request (HARQ) feedback acknowledgments to indicate the success of the downlink transmission, $2$) scheduling request to seek the time-frequency resources from cellular network for uplink transmission, and, $3$) downlink channel-state information (CSI) for link adaptation.

The importance of physical layer security motivates researchers to move in this direction. The method in \cite{ML} develops a feature-based classification
model using conventional machine learning (ML) algorithms. Similar works that use ML methods are not successful in case of insufficient training samples \cite{lastmm}, \cite{deeep}. \cite{mimooo} works on the passive jamming on downlink cell-free massive multiple-input multiple-output (MIMO) systems.
This work broadcasts artificial noise to disrupt the jamming signal. It proposes a cooperative physical layer security algorithm to facilitate artificial noise broadcasting via access point cooperation. Then, it presents an independent physical layer security algorithm to almost retake the consumed power using access points with independent artificial noise broadcasting. Obviously, the power consumption problem is the drawback of this method. The method in \cite{amm} is designed for only the particular adaptive modulation OFDM systems, and, finally, \cite{SINR} studies the maximization problem of signal-to-interference plus noise ratio (SINR), which is infirm against low power jamming attacks.

\begin{figure}[t]
\centering
        \includegraphics[width=2.5in]{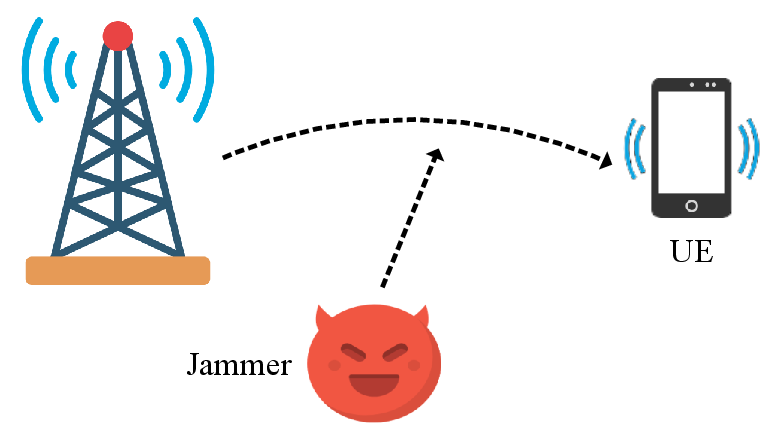}
     \caption{The considered system model.}
    \label{fig:2}
\end{figure}

OFDM is used on both uplink and downlink of $5$G NR standard and it seems to be an inseparable scheme for the third generation of mobile networks onwards. In this study, we investigate the loss of orthogonality (LoO) property in the $5$G NR physical channels, and then propose a method for jamming cancellation at the user side. The contributions of this study are as follows: $1)$ thanks to deploying the method at the user side, this work is independent of the number of jammers, $2)$ no additional hardware is required, and, $3)$ due to our focus on the frequency aspect, the power of jamming signal does not affect the performance of our method. The considered system model is shown in Fig. \ref{fig:2}.

\section{Jamming Principle}


In OFDM systems, LoO between subcarriers leads to inter-carrier interference (ICI), and further results in severe DoS for the users. In the following, we study an attacking scenario where a jammer shifts the frequencies of the subcarriers, causing misalignment of the spectral nulls. This phenomenon results in ICI.

$N$-point inverse fast Fourier transform (IFFT) at the transmitter side generates one OFDM symbol as follows: 
\begin{equation}
x_n\triangleq\frac{1}{N}\sum_{k=0}^{N-1}X_k e^{j n w_k},~~n={0,...,N-1},
\end{equation}
where $k$, $n$ and $w_k\triangleq{\frac{2\pi k}{N}}$ are the frequency index, time index, and the $k$th radian frequency component, respectively. $X_k\triangleq A_k e^{j \psi_k} \in \mathbb{C}$ is the transmitted symbol on the $k$th subcarrier (i.e., $k$th frequency component) with amplitude $A_k$ and phase $\psi_k$. Thus, $x_n$ consists of $N$ subcarriers (i.e., $N$ values for $A_k$, $\psi_k$ and $w_k$).

Let $O_{ki}\triangleq\frac{1}{N}\sum_{n=0}^{N-1} c_k c_i^{\ast}$. The orthogonality condition for two different subcarriers $c_k\triangleq X_k e^{j n w_k}$ and $c_i\triangleq X_i e^{j n w_i}$, $\ k\neq i$, is defined by the following equation:
\begin{equation}
\begin{cases}
  \text{Orthogonal} & \text{if $O_{ki}=0$}, \\
  \text{Non-orthogonal} & \text{otherwise.}
\end{cases}
\end{equation}
Since SSB is unencrypted during the initial access, the jammer can simply acquire the value of $N$. If ${\tilde{c_i}}$ denotes the attacked $c_i$ by a malicious jammer with the signal $a_{m}={X_m}e^{j n w_m}$, then $\tilde{O}_{ki}\triangleq\frac{1}{N}\sum_{n=0}^{N-1} c_k \tilde{c}_i^{\ast}$ yields:
\begin{multline}\label{ok0}
\frac{1}{N}\sum_{n=0}^{N-1}   X_k e^{j n w_k} \sum_{n=0}^{N-1} {X_i^{\ast}}{X_m^{\ast}} e^{j n w_{-{\tilde{i}}}}\\=\frac{1}{N}X_k X_i^{\ast} X_m^{\ast}\mathcal{H}_{k,i},
\end{multline}
where $\mathcal{H}_{k,i} \triangleq \sum_{n=0}^{N-1} e^{j n w_{k-{\tilde{i}}}}$ and ${\tilde{i}}\triangleq i+m$. Using the following equation:
\begin{equation}
\sum_{i=0}^{N-1} e^{-j d w_{i}}=
\begin{cases}
   N & \text{for $d=z\times N$, $z\in\mathbb{Z},$} \\
  0 & \text{otherwise,}
\end{cases}
\end{equation}
we have:
\begin{equation}
\tilde{O}_{ki}=
\begin{cases}\label{ok1}
  A_k\times A_i^\ast\times  A_m^{\ast} & \text{for $\frac{k-{\tilde{i}}}{N}=z\in\mathbb{Z},$} \\
  0 & \text{otherwise.}
\end{cases}
\end{equation}


\begin{algorithm}[t]\label{algo1}
\caption{The proposed anti-jamming scheme for OFDM-based physical channels.}
\textbf{Input:} $c_k\triangleq X_k e^{j n w_k}$, $c_i\triangleq X_i e^{j n w_i}$, $a_{m}\triangleq{X_m}e^{j n w_m}$ and $N$,
\begin{algorithmic}[1] 

\item \textbf{for}{ $i = 0$ to $N-1$,}

\State{Compute $\psi(i)\triangleq
\sum_{k=1}^{{N-1}}\mathbf{1}\{{\textbf{Tr}(\mathbf{A}_{k,i})\neq 0 \}}$},

\If{$\psi(i)=N-1$,} 
\State{{\textbf{Return:}} $i$th subcarrier is under attack,}
\EndIf
\item \textbf{end for}

\If {$m^{\prime}+k+{\tilde{i}}\neq zN$, \textbf{for}{~all $k\in\{0,..., N-1$\},}}
\State{{\textbf{Return:}} $m^{\prime}$,}
\Else ~~{{Update}} $m^{\prime}$,

\EndIf
\end{algorithmic}
\textbf{Output:} $c_{m^{\prime}}\triangleq X_{m^{\prime}} e^{j n w_{m^{\prime}}}$.
\end{algorithm}

\section{Anti-Jamming Strategy}

In this section, we detect the attacked subcarriers, the ones that have lost the orthogonality property, and then we propose multiplication of a new signal by the received signal at the user side, with the aim of jamming cancellation. Let $\mathbf{e}^m\triangleq \{e^{0}, e^{j w_{-m}},..., e^{j (N-1) w_{-m}}\}^{T}$, $\mathbf{e}^{k,i}\triangleq \{e^{0},e^{j w_{k-i}},..., e^{j (N-1) w_{k-i}}\}$, and $\mathbf{A}_{k,i}^{N\times N}\triangleq \mathbf{e}^{m} \times \mathbf{e}^{k,i}$. Referring to (\ref{ok0}) and (\ref{ok1}), $\mathcal{H}_{k,i}=\textbf{Tr}(\mathbf{A}_{k,i})$, and, if the $i${th} subcarrier is attacked by a jammer, then $\textbf{Tr}(\mathbf{A}_{k,i})\neq 0$.
Therefore, in order to determine the $i$th subcarrier that is under a jamming attack, we propose the following test:
\begin{equation}\label{test}
\psi(i)\triangleq
\sum_{k=1}^{{N-1}}\mathbf{1}\{{\textbf{Tr}(\mathbf{A}_{k,i})\neq 0 \}},
\end{equation}
where $\mathbf{1}\{.\}$ is the indicator function, and $i,k \in \{0,...,N-1\}$. Equation (\ref{test}) consists of $\tilde{N}\triangleq\frac{N\displaystyle !\,}{2\displaystyle !\,(N-2)\displaystyle !\,}$ non-repetitive executions to test each subcarrier with the other $N-1$ subcarriers. Therefore, we determine the $i$th disrupted subcarrier using the following statement:
\begin{equation}\label{4}
\begin{cases}
    LoO: & \text{$\psi(i)=N-1,$}\\
  No~fully~LoO: & \text{otherwise},
\end{cases}
\end{equation}
where no fully LoO refers to the situation in which alignment of all the spectral nulls is not entirely missed.

\begin{figure}[t]
\centering
        \includegraphics[width=3.5in]{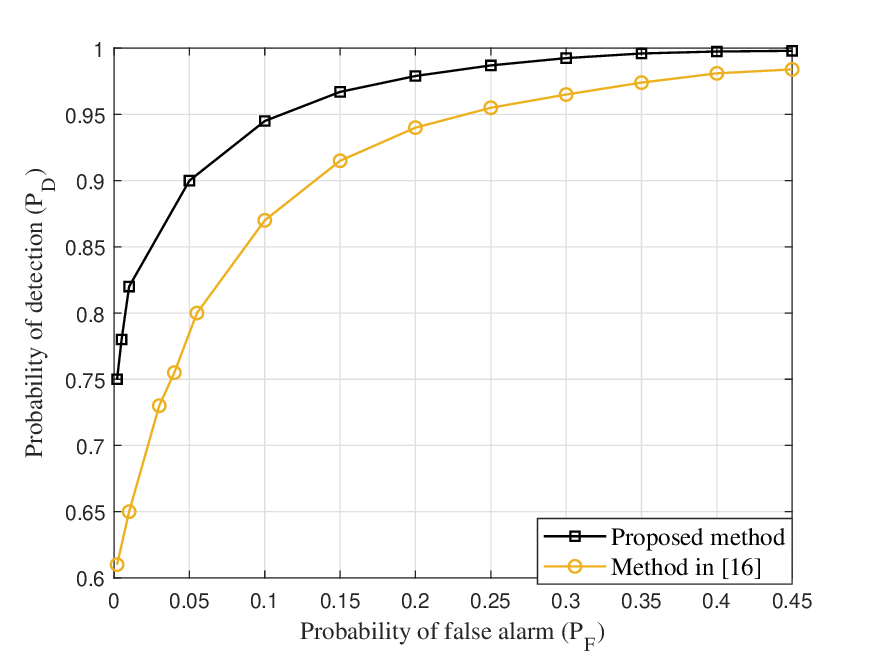}
     \caption{The detection performance of the proposed method compared with \cite{SINR} for various $N$ values.}
    \label{fig:3}
\end{figure}

\begin{remark}
Equation (\ref{ok1}) declares that the condition for the occurrence of LoO between $c_i$ and $c_k$ is $\frac{k-{\tilde{i}}}{N}=z\in\mathbb{Z}$. However, this condition (i.e., results in LoO) may caused by one or both of the following reasons: $1$) a jamming attack has occurred, $2$) the value of $N$ is not chosen correctly and needs to be updated.
\end{remark}


The anti-jamming approach is based on multiplication of signal $c_{m^{\prime}}\triangleq X_{m^{\prime}} e^{j n w_{m^{\prime}}}$ by the $i$th subcarrier such that $m^{\prime}\neq zN-(k+{\tilde{i}})$, for any $k \in \{0,...,N-1\}$ and $z\in\mathbb{Z}$. We summarize the design steps of our proposed method to detect and neutralize a jamming attack as the pseudo-code in Algorithm $1$.

\section{Simulation Results}
In this section, we illustrate the simulation results of the proposed detection method using receiver operating characteristic (ROC) curve. ROC is the plot of the detection probability $(P_D)$ versus the false alarm probability $(P_F)$ at each threshold setting. We consider clustered delay line (CDL)-D model with the parameters according to
$3$GPP $38.901$ \cite{ref13}, and subcarrier
spacing of $15$kHz. The UE and the jammer are uniformly
distributed into $(-2\pi,2\pi)$ and the BS is located at the
center of this circular region. Furthermore, jamming to signal ratio (JSR) and SNR are set to $0$dB and $5$dB, respectively.

Referring to Remark $1$, LoO is not only the result of a jamming attack. Fig. \ref{fig:3} depicts the superiority of the proposed method over the competing method in \cite{SINR}, in the aspect of detection performance. This figure is achieved using $N= 256,..., 2048$ to consider the effect of different subcarrier numbers. The future efforts in this direction will be conducted in a real-time system, to estimate all the attacked subcarriers, while only a few of them are available.




\section{Conclusion}\label{defin6}
In this study, we have investigated the LoO phenomenon in the $5$G NR physical channels, which is occurred by a malicious attacker. Firstly, we used the non-orthogonality condition to detect the attacked subcarriers at the user side, and then a multiplicative signal was proposed to neutralize the jamming attack. Simulation results evaluate our proposed method with the different number of subcarriers and prove its detection superiority over a recent method. Vulnerability of OFDM-based physical channels, specially PBCH and PDCCH, motivates researchers to further studies in this direction.

\leavevmode%


\ifCLASSOPTIONcaptionsoff
  \newpage
\fi
\bibliographystyle{IEEEtran}
\bibliography{keylatex}
\end{document}